\documentclass[preprint, aps, pre, eqsecnum,amsmath,amssymb,showpacs]{revtex4}              

\usepackage[final]{graphicx} 

\newcommand{\Nabla}{\mbox{\bf\boldmath $\nabla$}}                              
                                                             
\renewcommand{\vec}[1]{{\bf #1}}

\newcommand{\bean}{\begin{eqnarray}}
\newcommand{\eean}{\end{eqnarray}}
\newcommand{\bea}{\begin{eqnarray*}}
\newcommand{\eea}{\end{eqnarray*}}
\newcommand{\beq}{\begin{equation}}
\newcommand{\eeq}{\end{equation}}

\def\vereq#1#2{\lower3pt\vbox{\baselineskip1.5pt \lineskip1.5pt
\ialign{$\hfill##\hfil$\crcr#2\crcr\sim\crcr}}}

\begin{document}

\title{Oscillatory convection in binary mixtures: thermodiffusion, 
solutal buoyancy, and advection}

\author{D.~Jung, P.~Matura, and M.~L\"{u}cke}
\affiliation{Institut f\"{u}r Theoretische Physik, Universit\"{a}t
des Saarlandes, D-66041~Saarbr\"{u}cken, Germany\\}

\date{\today}

\begin{abstract}

The role of thermodiffusive generation of concentration fluctuations via the Soret
effect, their contribution to the buoyancy forces that drive convection, the 
advective mixing effect of the latter, and the diffusive homogenisation are
compared and elucidated for oscillatory convection. Numerically obtained solutions
of the field equations in the form of spatially extended 
relaxed traveling waves, of standing waves, and of the transient growth of standing
waves and their transition to traveling waves are discussed as well as spatially
localized convective states of traveling waves that are surrounded by the quiescent
fluid.
\end{abstract}

\pacs{47.20.-k, 47.20.Ky, 47.54.+r, 47.27.Te }

\maketitle

 \section{Introduction}
 \label{Sec:Intro}
 Convection in binary miscible fluids like ethanol-water, $^3$He 
$-^4$He, or various gas mixtures is an example of systems that show rich 
and interesting pattern formation behavior when driven away from 
equilibrium. It is paradigmatic for problems related to
instabilities, bifurcations, 
selforganisation, complex spatiotemporal behavior, and turbulence.
Compared to convection in one-component 
fluids the spatiotemporal properties are far more 
complex. The reason is that concentration variations 
which are generated via thermodiffusion, i.e., the Soret effect  by 
externally imposed and by internal temperature gradients influence the buoyancy,
i.e., the driving force for convective flow. The latter in turn mixes by
advectively redistributing concentration. This nonlinear advection
gets in developed convective flow typically much larger than the smoothening 
by linear diffusion --- P\'eclet numbers measuring the strength
of advective concentration transport relative to diffusion are easily of the
order thousand. Thus, the concentration balance is strongly nonlinear 
giving rise to strong variations of the concentration field and to
boundary layer behavior. In contrast to that, momentum and 
heat balances remain weakly nonlinear close to onset as in pure fluids implying
only smooth and basically harmonic variations of velocity and temperature fields
as of the critical modes. Without the thermodiffusive Soret coupling between 
temperature and
concentration field initial concentration deviations from the mean diffuse away and
influence no longer the balances of the other fields [such a system displays 
close to onset only the familiar stationary Benard rolls that are called
for historical reasons states of stationary overturning convection (SOC)].

Hence, the feedback interplay between ({\em i}) the Soret generated 
concentration variations, ({\em ii}) the resulting modified buoyancy, and
({\em iii}) the 
strongly nonlinear advective transport and mixing causes binary mixture 
convection to be rather complex with respect to its spatiotemporal
properties and its bifurcation behavior.

Take for example the case of negative Soret coupling, $\psi < 0$, 
between temperature and concentration fields \cite{coupling}. Then the above 
described feedback interplay generates oscillations. In fact the buoyancy 
difference in regions
with different concentrations was identified already in \cite{WKPS85} as the 
cause for traveling wave convection.

Oscillatory convection appears in the form of the
transient growth of convection at supercritical heating, in spatially
extended relaxed 
nonlinear traveling wave (TW) and standing wave (SW) solutions that branch 
subcritically
out of the conductive state via a common Hopf bifurcation, and in spatially
localized traveling wave (LTW) states. TW and LTW convection has been studied
experimentally and theoretically in detail 
\cite{review,Ahlers,Behringer,Kolodner,Platten,Steinberg,Surko,Yahata,Knobloch,JL02}.
The transient oscillatory growth of convection was investigated by
numerical simulations \cite{FL}. Nonlinear SW solutions were obtained only 
recently \cite{MDL}. 

Here we compare these different oscillatory solutions and elucidate
common features and differences. 

In Sec.~\ref{Sec:System} we describe the system
and our numerical methods for investigating it. Sec.~\ref{Sec:extended} deals with
extended states of TWs, transient growth of SWs, with SW $\to$ TW transitions, and
with relaxed SWs. Then in Sec.~\ref{Sec:LTW} LTWs are discussed. 
Sec.~\ref{Sec:conclusion} contains a short conclusion.

 \section{System}
 \label{Sec:System}
 
  We consider a binary fluid layer of thickness $d$ at a mean temperature 
  $\bar{T}$ 
 and with a mean concentration $\bar{C}$ of the lighter component (e.g. ethanol
  in water). 
 It is confined between two solid horizontal plates perpendicular to
 the gravitational acceleration $g$. 
 The layer is heated from below with a temperature difference $\Delta T$ 
 to the upper plate.
 The variation of the fluid density $\rho$ due to temperature and
 concentration variations is governed by the linear thermal and solutal
 expansion coefficients
 $ \alpha = - \frac{1}{\rho}\frac{\partial\rho}{\partial \bar{T}} $ and 
 $ \beta = - \frac{1}{\rho}\frac{\partial\rho}{\partial \bar{C}} $,
 respectively. Both are positive for ethanol-water.  The solutal diffusivity 
 of the binary  mixture is $D$, its thermal diffusivity is $\kappa$, and its 
 viscosity is $\nu$. 

 The vertical thermal diffusion time $d^2/\kappa$ is used as
  the time scale of the system and all velocities are
  scaled by $\kappa/d$. 
 Temperatures are reduced by the vertical temperature
  difference $\Delta T$ across the layer and concentration deviations from 
  the mean concentration by $\frac{\alpha}{\beta}\Delta T$.
 The scale for the pressure is given by $\frac{\rho\kappa^2}{d^2}$.
 Then, the balance equations for mass, momentum, heat, and concentration
  \cite{LL66,PL84} read in Oberbeck--Boussinesq approximation \cite{BLKS95I}
 \begin{mathletters}
 \label{eq:baleqs}
 \begin{eqnarray}
 \Nabla \cdot \vec{u} = 0 \label{eq:baleqmass}\\
 \partial_t\, {\bf u}   =    - {\bf \mbox{\boldmath $\nabla$} }\,
({\bf u : u}  +  p
- \sigma \, {\bf \mbox{\boldmath $\nabla :  $}\,\, u }) +
{\bf B} \,\,\,;\,\,\,  
{\bf B}    =    \sigma\, R\, (\delta T + \delta C) {\bf e}_z \label{eq:baleqveloc}\\ 
 \partial_t\delta T  = - \Nabla \cdot \vec{Q}\ =\ 
  - \Nabla \cdot \left[ \vec{u}\delta T - \Nabla \delta T\right]\label{eq:baleqheat}\\
 \partial_t \delta C  =  - \Nabla \cdot \vec{J}\ =\
  - \Nabla \cdot \left[ \vec{u} \delta C - L\Nabla\left(\delta C -\psi \delta T\right) \right]\ .
  \label{eq:baleqconc}
 \end{eqnarray}
 \end{mathletters}
 Here, $\delta T$ and $\delta C$ denote deviations of
  the temperature and concentration fields, respectively, from their global
  mean values $\bar{T}$ and $\bar{C}$; $\vec{Q}$ and
  $\vec{J}$ are the associated currents. $\bf B$ is the buoyancy.
 The Dufour effect \cite{HLL92,HL95}
  that provides a coupling of concentration gradients into
  the heat current $\vec{Q}$ and a change of the thermal diffusivity
  is discarded in (\ref{eq:baleqheat}) since it is relevant
  only in few binary gas mixtures \cite{LA96} and possibly in
  liquids near the liquid--vapor critical point \cite{LLT83}.

 Besides the Rayleigh number $R=\frac{\alpha g d^3}{\nu \kappa}\Delta T$
  measuring the thermal driving of the fluid three additional
  numbers enter into the field equations: the Prandtl number
  $\sigma=\nu/\kappa$, the Lewis number $L=D/\kappa$, and the separation ratio 
  $\psi=-\frac{\beta}{\alpha}\frac{k_T}{\bar{T}}= 
  - S_T\bar{C}(1-\bar{C})\frac{\beta}{\alpha}$. 
  Here $k_T = \bar{T}\bar{C}(1-\bar{C})S_T$ is 
  the thermodiffusion coefficient \cite{LL66} and $S_T$ the Soret coefficient. They
  measure changes
  of concentration fluctuations due to temperature gradients in the fluid.
$\psi$ characterizes the sign and the strength of the   Soret effect. 
 Negative Soret coupling $\psi$ (i.e., positive $S_T$ for mixture like ethanol
 water with positive $\alpha$ and $\beta$)
 induces concentration gradients of the lighter component
 that are antiparallel to temperature gradients. 
 In this situation, the buoyancy
  induced by solutal changes in density is opposed to the thermal buoyancy.
 Throughout this paper we consider mixtures with $L=0.01$, $\sigma=10$, 
 and various negative values of $\psi$ that are easily accessible with 
 ethanol-water experiments. 

 When the total buoyancy exceeds a threshold, convection sets in --- typically
  in the form of straight rolls. For sufficiently negative $\psi$ the primary 
  instability is oscillatory \cite{osc-inst}. 
 Ignoring field
  variations along the roll axes we describe here 2D convection in an
  $x$--$z$ plane perpendicular to the roll axes with a velocity field
 \begin{equation}
 \vec{u}(x,z,t) = u(x,z,t)\,\vec{e}_x + w(x,z,t)\,\vec{e}_z \, .
 \end{equation}
 This 2D type of convection is
  commonly enforced experimentally in convection channels of small extension in 
  $y$-direction since the rolls are oriented preferentially perpendicular
  to the channel walls \cite{CH93}.

 To find the time-dependent solutions  of the partial differential equations
  describing convection we performed 
  numerical simulations with a modification of the SOLA code that is based on 
  the MAC method \cite{MAC-SOLA,PT83}. 
 This is a finite-difference method
  of second order in space formulated on staggered grids for the different 
  fields. The Poisson equation for the pressure field that results from taking the 
  divergence of (\ref{eq:baleqveloc}) was solved iteratively  using the
  artificial viscosity method \cite{PT83}. In addition we used multi-mode 
  Galerkin expansion methods \cite{HL02}. 
  
 The boundary conditions for the fields were as follows: 
  we used realistic no slip conditions 
 for the top and bottom plates at $z=\pm 1/2$
 and we assumed perfect heat conducting plates.
 Furthermore, impermeability of the horizontal boundaries for the 
 concentration was guaranteed by enforcing
 the vertical concentration current $\vec{e}_z\cdot\vec{J}$ to vanish at 
 both plates 
 \begin{equation}
 \label{eq:Cbound}
 \vec{e}_z\cdot\vec{J} =
  - L\partial_z\left(\delta C-\psi \delta T\right)(x,z=\pm 1/2;t) = 0\ \ \ .
 \end{equation}
  Laterally periodic extended convection 
  structures were simulated by imposing lateral perodicity in 
 $x$-direction with periodicity length $\lambda=2\pi/k$. Spatially localized
 convection states were obtained in domains of size up to 160.
To measure the strength of the thermal driving
  we use the reduced Rayleigh number $r = R/R_c^0$ 
  that is scaled by the critical Rayleigh number $R_c^0$ for onset
  of pure-fluid convection with the critical wave number $k_c^0$\,. 
 The analytical values are $R_c^0 = 1707.762$ and $k_c^0 = 3.11632$. 
 However, to enable comparison 
 with experimental, analytical, or numerical results we scale $R$ by the 
  threshold $R_c^0$ in question, e.g., of the finite differences numerical code
 \cite{CJT97}.  
 
To characterize convection 
  we monitored among others the evolution of the spatial
  maximum of the vertical velocity field $w_{\rm max}$, of the oscillation 
  frequency, and of the reduced variance 
 \begin{equation}
 \label{eq:Mdef}
  M = \sqrt{\langle \delta C^2\rangle/\langle \delta C^2_{\rm cond}\rangle}
 \end{equation}
  of the concentration field.
 Note that $0\leq M \leq 1$ measures to which extent the binary fluid is mixed. 
 The better it is mixed the smaller is the spatial mean 
 $\langle \delta C^2\rangle$ of the square of the concentration deviation 
  from its mean -- for spatially localized convection states $M$ is defined
  slightly differently, cf. Sec.~\ref{Sec:LTW}. 
 The boundary conditions, however, sustain always small Soret-induced
  concentration gradients against the action of advection and diffusion and 
  prevents $M$ from vanishing completely.
 On the other hand, in   the conductive state (denoted by the subscript 
 ``cond'') with a Soret-induced vertical concentration
 stratification given by $\delta C_{\rm cond}(z) = -\psi z$, there is no 
 advective mixing.  Its concentration variance, 
 $\langle \delta C^2_{\rm cond}\rangle = \psi^2/12$, is largest yielding 
  $M_{\rm cond}=1$.

 \section{Spatially extended oscillatory convection}
 \label{Sec:extended}
 
 The complexity of (oscillatory) convection in binary mixtures has been the 
 subject of many studies \cite{reviewCF} that are too numerous to be discussed
 here. A recent survey on papers adressing the role of the concentration 
 field may be found in the introduction of \cite{FL}.
 
In this section we compare characteristic properties of spatially extended 
 oscillatory convection in the form of relaxed TWs \cite{BLKS95I,LBBFHJ98,HL98},
 SWs \cite{MDL} and of
 oscillatory transients \cite{FL} into, say, a nonlinear TW.  
 In Fig.~\ref{Fig:BifUebersicht} we show the bifurcation diagrams 
 of maximal vertical flow velocity 
  $w_{max}$ and oscillation frequency $\omega$ versus reduced 
  Rayleigh number $r$ for the representative Soret coupling $\psi=-0.25.$

 At $r_{osc}=1.335$ a SW solution 
 as well as a pair of symmetry degenerate left and right 
 traveling wave solutions bifurcate out 
 of the conductive state with wave number $k=\pi$. At this Rayleigh number  
  the system shows a subcritical Hopf bifurcation with a Hopf 
  frequency $\omega_H(k=\pi, r_{osc})=11.23$. Here two facts are worth 
  mentioning: (i) the bifurcation threshold and the frequency
  are practically the same as for the critical waves with the critical wave number 
  $k_c=3.135$ and 
 (ii) TWs with wavelength $\lambda \simeq 2$ are often 
  observed in large-scale experimental setups. 
 This value corresponds to the minimum of $r_{osc}(k)$.
 Fig.~\ref{Fig:TWonsetsaddles} shows $r_{osc}(k)$ for different $\psi$. 
 The respective onset values for the phase velocities $v_p^H(k)=\omega_H(k)/k$
 depend only 
 weakly on the wave number $k$. The curves $v_p^H(k)$ are shifted to higher 
 values with increasing
 negative $\psi$ due to the higher Soret-induced concentration contrast in the 
 conductive state.
 The saddle node lines $r_s^{TW}(k)$ are more or less downwards shifted versions of 
 $r_{osc}(k)$.
 The size of the interval, $r_{osc}(k) - r_s^{TW}(k)$, of subcritical TWs
 increases when $\psi$ becomes more negative. 
 For the parameters of Fig.~\ref{Fig:TWonsetsaddles}
 the TW branches end at large amplitudes in a well mixed state of vanishing frequency
 by merging with the stationary SOC solution branch of the same wavelength.
 For the parameters of Fig.~\ref{Fig:BifUebersicht} this happens at $r^{\ast}=1.65$ 
 where the SOC solution branch is included for convenience as well.

 In contrast to the TW solutions, 
 the SWs remain unstable at least against a transition into a (transient) TW. 
 Their solution branches (cf. Sec.~\ref{Sec:SW}) end in period dubling cascades 
 \cite{MDL}.
 
 \subsection{Advective mixing and diffusion of concentration
 fluctuations}
 \label{Sec:Mixdiff}

\begin{figure}
\includegraphics[clip=true,angle=0,width=12cm]{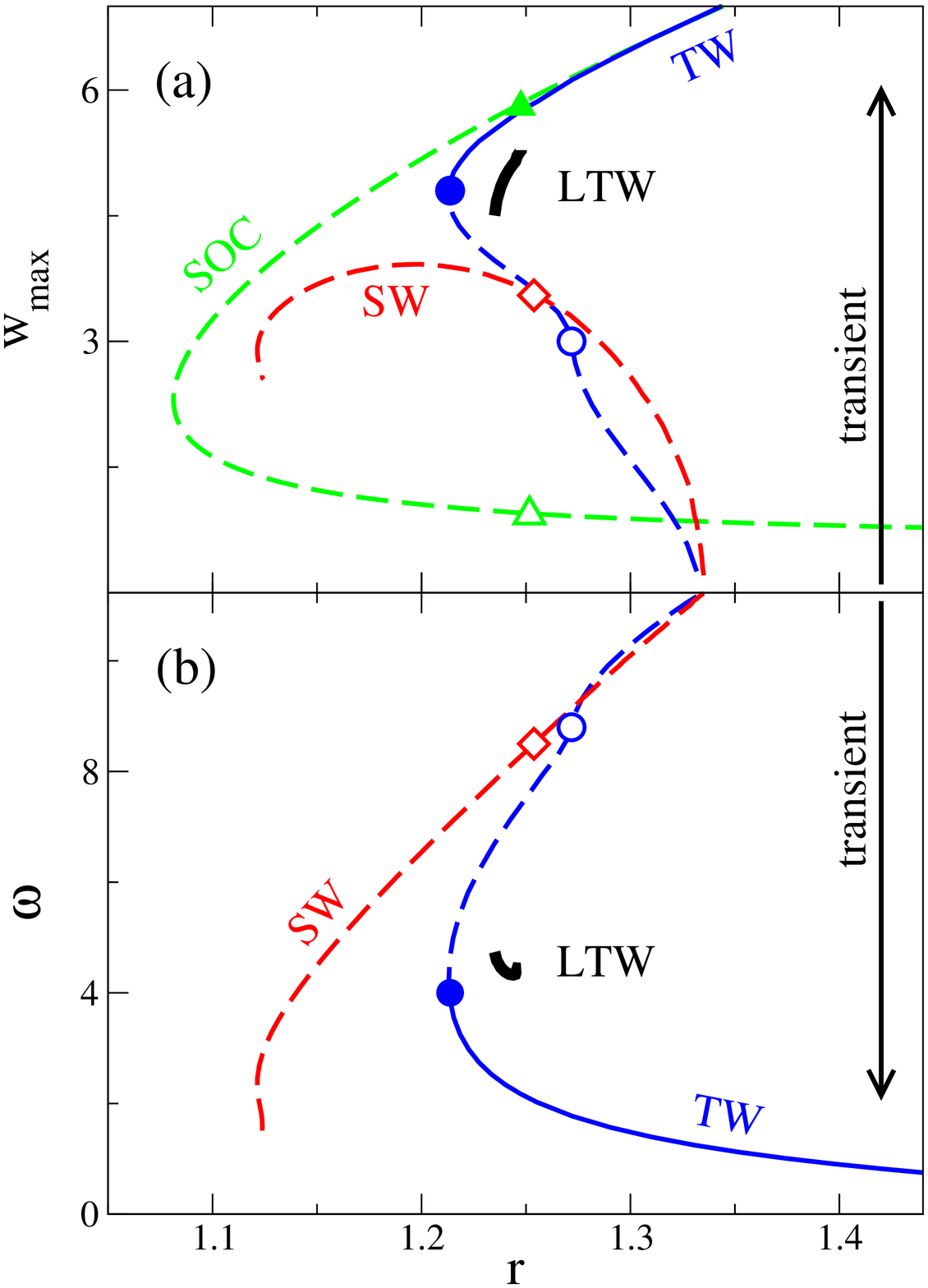} 
\caption{Bifurcation properties of maximal vertical flow velocity $w_{max}$
(a) and frequency $\omega$ (b) versus $r$ for representative relaxed nonlinear 
convective SOC, TW, SW, and LTW states. Full (dashed) lines refer to 
stable (unstable) states. Symbols identify the states that are discussed in 
more detail. TW and SW solutions bifurcate subcritically out of the conductive 
state at the oscillatory threshold $r_{osc}$ with common Hopf frequency. 
Vertical arrows indicate the supercritical transient of 
Fig.~\ref{Fig:transient}
from a growing SW perturbation of the quiescent conductive state into a 
nonlinear TW. The wavelength of SOC, TW, and SW is $\lambda = 2$. 
Parameters are $L = 0.01, \sigma = 10, \psi = -0.25$.}
\label{Fig:BifUebersicht}
\end{figure}

\begin{figure*}
\includegraphics[clip=true,width=12cm,angle=0]{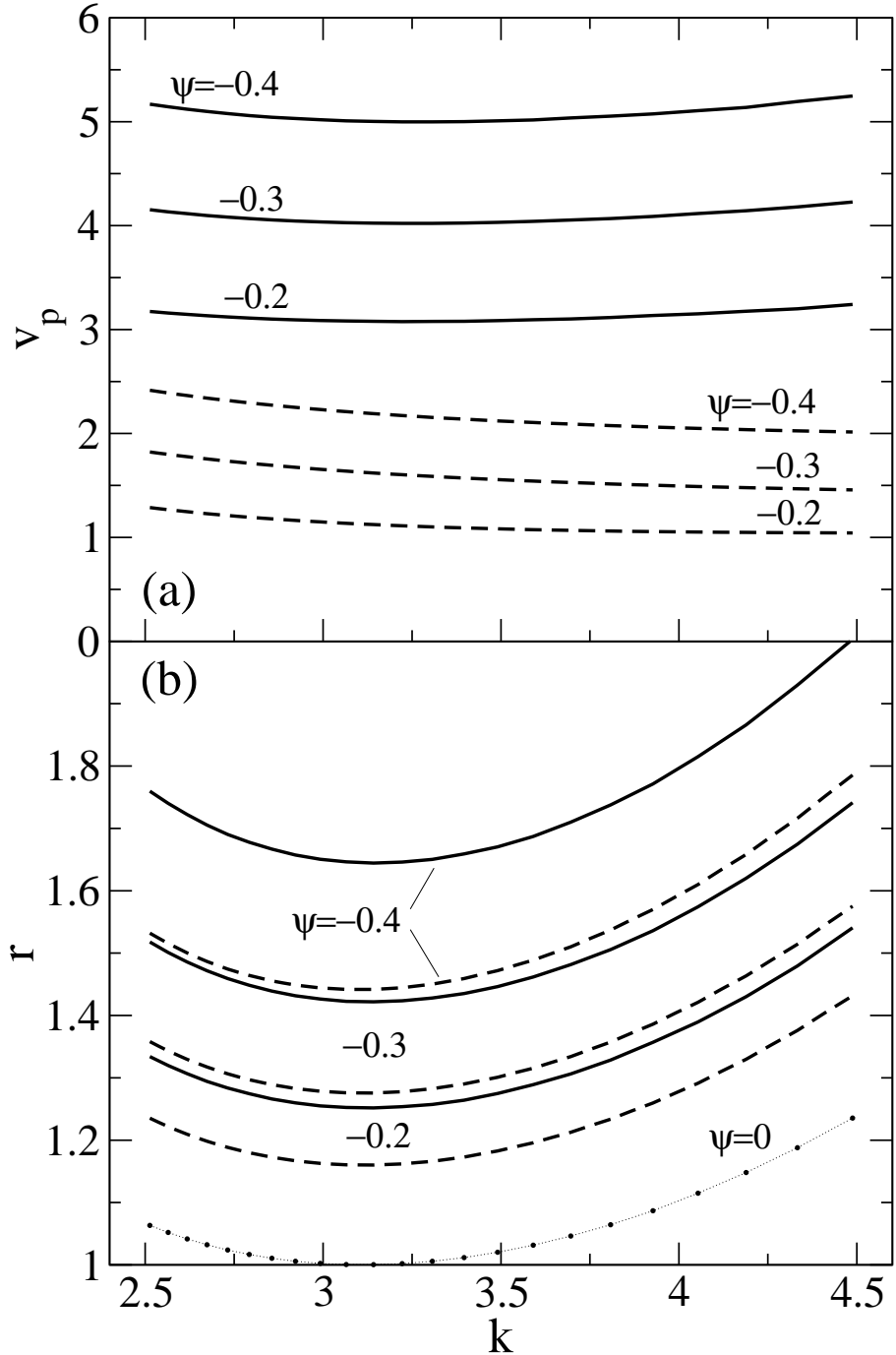}
\caption{Wave number dependence of TWs at onset and at their saddle-nodes
 for different $\psi$.
(a) Phase velocities $v_p(k)$ at the onset (straight lines) and at the 
saddle (dashed lines). 
(b) Location of the onset $r_{osc}(k)$ (straight lines) and of the saddle-nodes 
$r_{s}^{TW}(k)$  (dashed lines) for TW branches in the $k-r$-plane.
 The dotted line marks the onset of SOCs in pure fluids.
Parameters are $L =0.01$, $\sigma = 10$.
\label{Fig:TWonsetsaddles}}
\end{figure*}

\begin{figure}
\includegraphics[clip=true,width=15cm,angle=0]{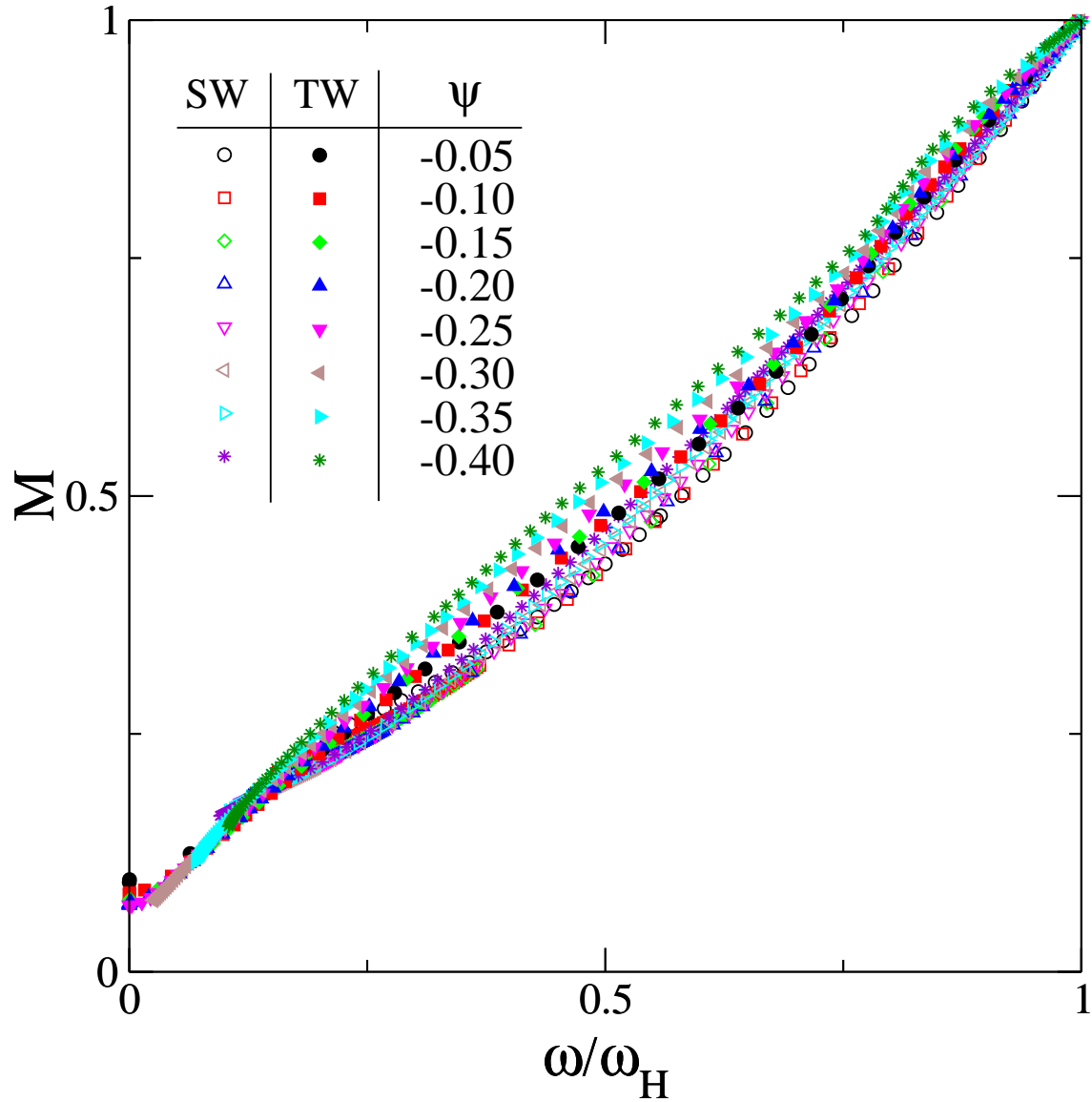}
\caption{Mixing number $M$ versus reduced oscillation frequencies of TWs and SWs
for various Soret coupling $\psi$. For SWs the time averaged mixing number is
plotted. Here $\omega_H$ is the respective Hopf 
frequency at onset. Parameters are $ \lambda = 2, L = 0.01, \sigma = 10$.  
\label{Fig:Mvsom}}
\end{figure}

 When moving along the 
 TW and SW solution branches in Fig.~\ref{Fig:BifUebersicht} the frequency 
 $\omega$ and the 
 TW phase velocity $v_p=\omega/k$ decrease monotonically starting from the large 
 Hopf value at the bifurcation threshold. This is related to the advective
 reduction of the Soret generated concentration gradients. 
 In fact there is  a  universal linear relation between the
 degree of mixing and the frequency $\omega$ which holds for weakly nonlinear as
 well as for strongly nonlinear states and for transients alike. 
 See for example \cite{HBL97,LBBFHJ98} for TWs, \cite{JL02} for LTWs, and \cite{FL}
 for transients.  This scaling relation $\omega/\omega_H \simeq M$ holds also for 
 SWs as one can infer from Fig.~\ref{Fig:Mvsom}. Also TWs with wavelength 
 different from 2 show roughly this scaling. 
  
\begin{figure}
\includegraphics[clip=true,width=8.5cm,angle=-90]{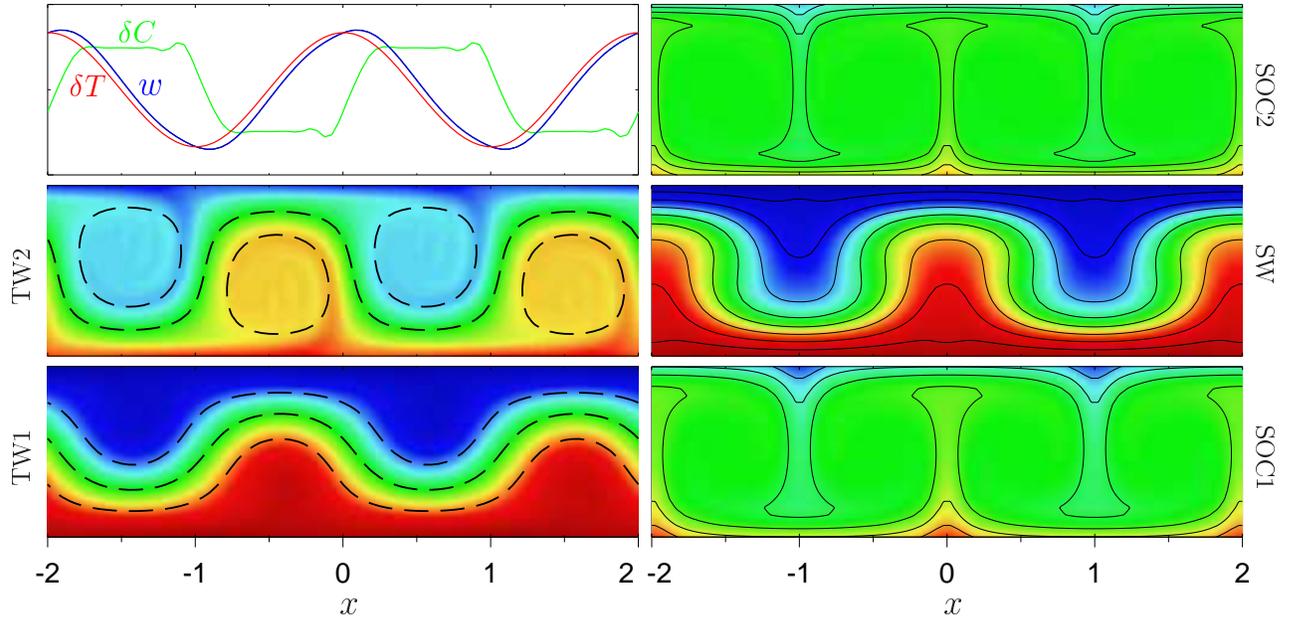}
\caption{Concentration distribution of selected extended states in vertical
cross-sections through the rolls. TW1(2) and SOC1(2) are
states  marked by open (filled) circles and triangles, respectively, 
in Fig.~\ref{Fig:BifUebersicht}. The SW refers to the lozenge in 
Fig.~\ref{Fig:BifUebersicht}. The top figure in the left column shows the
lateral wave profiles of $\delta C,\delta T$, and $w$ at midheight, $z=0$, of
the fluid layer in the TW2.
Parameters are $ \psi = -0.25, L = 0.01, \sigma =  10, \lambda = 2$.  
\label{Fig:comparestruct}}
\end{figure}

  In particular on the stable, strongly nonlinear 
 TW solution branch the binary fluid gets 
 with increasing $w_{max}$ and thermal
 driving $r$ more and more mixed and $M$ reduces almost to zero -- only  
 concentration gradients in the narrow boundary layers survive. This holds even
 more so for the well
 mixed SOC states (see e.g. the SOC2 in Fig.~\ref{Fig:comparestruct} which is 
 marked in Fig.~\ref{Fig:BifUebersicht} by the filled
 triangle) that resemble closely the corresponding stationary
 state in the pure fluid with the same $\sigma$ and $r$.

 The transition to convection at
  $r_{osc}$ is hysteretic, i.e., of first order because the Soret coupling
  coefficient $\psi=-0.25$ is sufficiently negative.   
 The associated precipitous growth of 
  convection is caused by an interplay 
  between the solutal contribution to the buoyancy 
  \cite{BLKS95I,LBBFHJ98} that tends to stabilize the conductive state and the
  effect of advective mixing. The latter enhances convection by reducing the 
  adverse effect of the Soret generated concentration variations.
  
  There occurs an ``S''-shaped deformation of the unstable part 
 of the TW bifurcation branch [dashed (blue) line in Fig.~\ref{Fig:BifUebersicht}]
 when the advection velocity $w_{max}$ has grown to become equal 
 to the TW phase velocity $v_p$.  Then the first
 closed streamlines appear \cite{HL98} in the frame of reference
 that is comoving with the TW phase velocity and where 
 the TW solution is time independent. 
 This is indeed a characteristic feature of strongly nonlinear TW convection, cf.
 further below. 
 
 For $w_{max}<|v_p|$, i.e.,
 closer to the threshold $r_{osc}$ all streamlines are open and the field structures
 look almost harmonic -- see the TW1 plot in Fig.~\ref{Fig:comparestruct}. 
 This weakly nonlinear TW, marked by an open circle in 
 Fig.~\ref{Fig:BifUebersicht}, is just at the border  $w_{max} \simeq |v_p|$. 
 This condition marks also the border line beyond which a straightforward 
 small-amplitude expansion around the convective onset breaks down \cite{HBL97}.
 In addition the generic {\em transient growth} dynamics of 
 oscillatory convection (consisting initially of oppositely traveling waves of 
 roughly equal velocity amplitudes $|A_R| \simeq |A_L|$) 
 undergoes a dramatic change that cannot be described at all with 
 amplitude equation models  when the flow amplitudes $|A_{R,L}|$ approach
 the border line value of $\omega/k$ \cite{FL}. 

 The qualitative change in the flow topology between weakly and strongly nonlinear
 TWs causes a different mixing behavior with increasing amplitude. This is the 
 reason for the
 ``S''-shaped deformation of the unstable part 
 of the TW bifurcation branch [dashed (blue) line in Fig.~\ref{Fig:BifUebersicht}].
 It becomes more pronounced with $\psi$ becoming more negative (see e.g. 
 Fig.~\ref{Fig:BifTW_LTW}). Eventually the strongly nonlinear TW solution branch 
 develops a
 bistable part there on which fast TWs are located \cite{HBL97,LBBFHJ98}.
 
 For relaxed TWs with $0<|v_p|<w_{max}$ the remaining open streamlines are 
 spatially correlated 
 with inner boundary layers of the concentration field. 
 As indicated in the TW2 of Fig.~\ref{Fig:comparestruct} they meander along the 
 (green) regions of mean concentration,
 $\delta C = 0$, between and around the roll-like regions of closed streamlines.
 In a right-propagating TW the regions of
 closed streamlines for the right (left) turning fluid domains are 
 rich (poor) in the lighter component - here ethanol - and they are 
 displaced towards the upper cold (lower
 warm) plate, where the Soret effect maintains a boundary layer with
 alcohol surplus (deficiency). In addition the meandering open streamlines lie
 between the upper (lower) closed streamline regions and the opposite bottom (top)
 concentration boundary layer. This structure of closed and open streamlines causes
 the top (bottom) boundary layer to feed high (low) concentration 
 only into the right (left) turning roll domain at the location of 
 downflow (upflow). Then the fluid becomes 
  diffusively homogenized in the closed streamline regions of the rolls 
  leading to anharmonic concentration 
  profiles of trapezoidal shape shown in the top plot of the left column of
  Fig.~\ref{Fig:comparestruct}.
 The motion of the rolls with their specific concentration distribution implies 
  a mean concentration current 
 which is directed to the right in the upper and to the left in the lower half of
 the layer. See refs. \cite{BLHK89,BLK92,BLKS95I} for a more detailed discussion 
 of this current.
 
With increasing $w_{max}$ and decreasing $|v_p|$
the regions of closed streamlines grow at the expense of the open ones. Thereby,
the former also come closer to the respective opposing boundary layer. This
decreases the asymmetry of the boundary layer
 feeding into oppositely turning rolls. As a consequence the 
 concentration contrast between adjacent TW rolls decreases until in the SOC 
 state with $v_p=0$ the rolls are fed symmetrically by both boundary layers and 
 mirror symmetry between the rolls is established. 
 
 \subsection{Symmetries}
 \label{Sec:Symmetries}

 SOCs and SWs are laterally mirror symmetric around positions
of maximal up- and downflow, e.g. $x$=0 in Fig.~\ref{Fig:comparestruct}, 
and the node locations of the fields
are fixed in time. This symmetry is broken in TWs -- see, e.g., the 
concentration
contrast between left and right turning rolls in Fig.~\ref{Fig:comparestruct}. 
But all fields of SOCs, SWs, and TWs have at every
instant definite parity under the mirror-glide (MG) operation $(x,z)\to
(x+\lambda/2,-z)$ of vertical reflection at mid-height, $z$=0, combined with 
lateral translation by half a wavelength. We did not observe SWs without
this symmetry -- perturbations breaking it that we introduced for test
purposes always decayed rapidly to zero. 

Furthermore, all transients investigated in \cite{FL} obeyed the MG symmetry 
with the exception of the very early stage in cases where the imposed initial 
conditions were not MG symmetric.  But even then
  the MG symmetry was rapidly restored by a fast decay of MG symmetry-breaking
  modes. Also the transient growth seen in the experiments in an annular 
  geometry by Winkler and Kolodner \cite{WK92} was locally MG symmetric. 
 A time dependent generalization
  of this symmetry was found to be realized in LTW
  states \cite{LTW} and an extension to 3D patterns was observed in 
  \cite{JHL98}. Furthermore, MG symmetric convective structures 
  were not only observed in Soret
  driven convection with only temperature gradients imposed but also in 
  thermosolutal convection \cite{STK98}. 

 Thus, the MG symmetry that is displayed by the basic conductive state 
  and by the linear critical convective modes seems to 
  be quite robust and also persistent in non-linear 
  convective structures of pure fluids and of mixtures. 
 The robustness of this symmetry is remarkable 
  given that the nonlinearity in the concentration balance, i.e., 
  the Peclet number $w/L$ is quite large -- of the order of 1000.   
 However, the concentration field is ``tamed'' by being coupled to
  the velocity and temperature fields. Their shape
  remains mostly harmonic for small
  supercritical thermal driving like in pure fluid convection. And thus the 
  increase in structural 
  complexity associated with a MG symmetry breaking does not occur.

Finally, the SW fields start at onset to have in addition a 
definite  mirror-timeshift symmetry (MTS), e.g., $f(x,z,t)=-f(x,-z,t+\tau/2)$ 
for $f=\delta C,\delta T$, and the vertical velocity field $w$ with
$\tau=2\pi/\omega$ being the SW oscillation period. At mid-height the
condition $f(t)=-f(t+\tau/2)$ implies in particular that positive and negative
field extrema of an oscillation cycle have equal magnitudes.  
SWs with smaller frequency that are located in the end region of the SW solution
branches break this symmetry (cf. Sec.~\ref{Sec:SW}). This is a prerequisite for 
period doubling \cite{SW84}.

 \subsection{Supercritical growth and SW$\to$TW transition}
 \label{Sec:transient}
Here we review the supercritical spatiotemporal behavior of SW transients.
We start from the quiescent conductive state at $r<r_{osc}$, disturb it 
slightly by adding small random numbers in the range of 
$\left[ -10^{-4}, 10^{-4}\right]$,  and 
 simultaneously increase the control 
parameter $r$ slightly above threshold as indicated in Fig.~\ref{Fig:BifUebersicht}.
 Then one observes a 
generic transition scenario that occurs similarly also in analogous
experiments. It consists of three evolution phases that we have found to be
 generic for convective growth out of small unspecific perturbations: 
({\it i}) an exponentially growing SW of high frequency over a time interval 
the  length of which depends on the size of the initial perturbations of the 
conductive  state, ({\it ii}) an intermediate phase that is always very short 
with a  spatiotemporally 
 complicated transformation from SW into a high-frequency TW and finally 
 ({\it iii}) a long-term TW transient to a low-frequency, 
 strongly nonlinear, relaxed TW or SOC state depending on $r$ \cite{FL}.
 
\begin{figure}
\includegraphics[clip=true,width=15cm,angle=0]{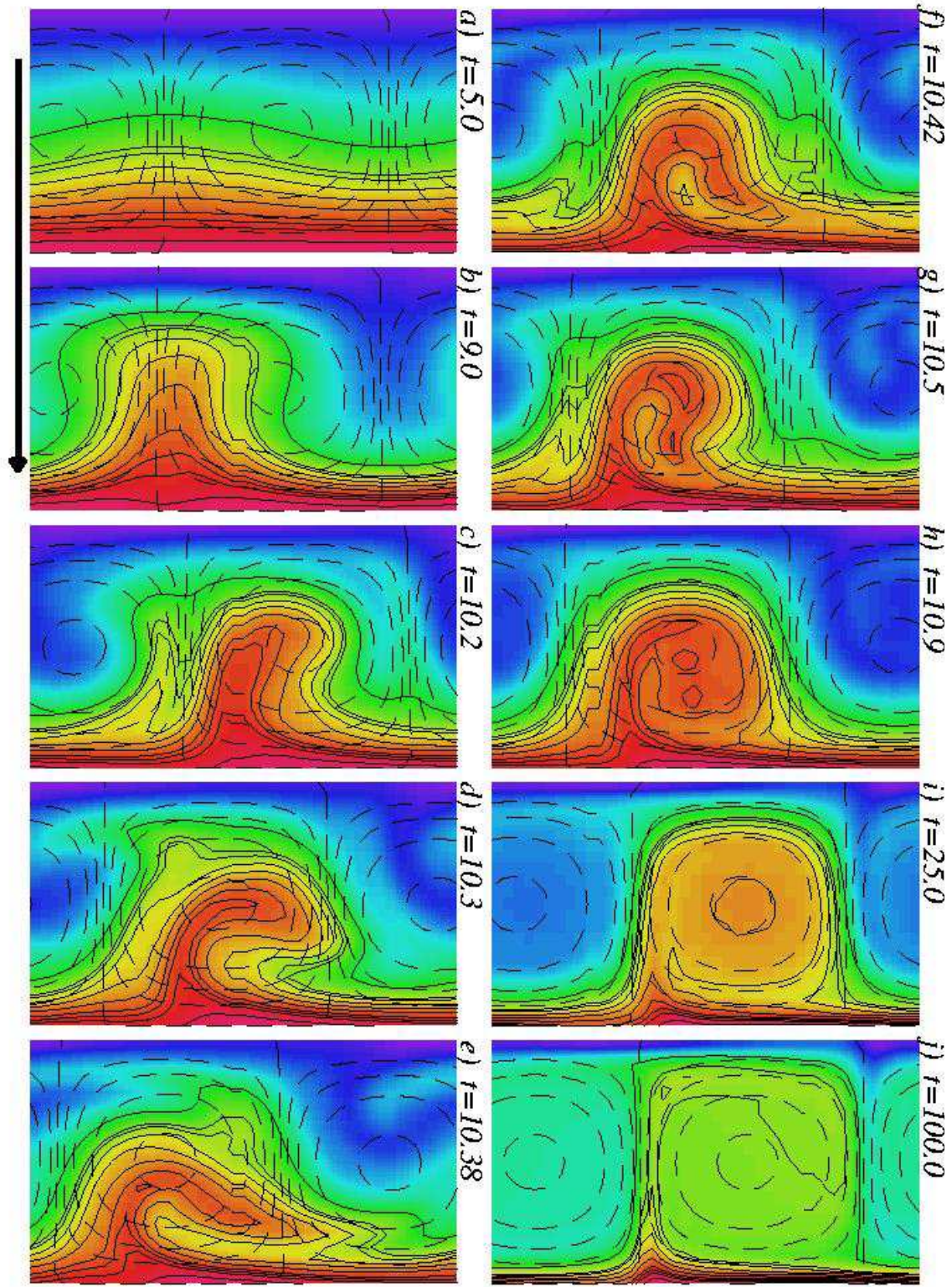}
\caption{Time ordered snapshots of the concentration redistribution in the SW$\to$TW
transformation. The concentration distribution in a vertical cross section
of the fluid layer is displayed for one wavelength. 
Full isoconcentration lines  are shown for $\delta C>0$. Dashed lines are 
streamlines, i.e., tangents to
the instantaneous velocity field. The final TW (j) propagates to the left. 
Parameters are 
$L =0.01$, $\sigma = 10$, $\psi = -0.25$, $r = 1.42$, and $\lambda = 2$.}
\label{Fig:transient}
\end{figure}

In generic, i.e., non specific initial perturbations that break the 
mirror symmetry 
$x \to - x$ only weakly the two critical oscillatory modes of right and left
traveling waves that can grow above the Hopf
threshold are contained with roughly equal 
amplitudes, i.e., $|A_R| \simeq |A_L|$.
Thus, initially, i.e., as long as linear theory applies, the two critical TW modes 
that started with $|A_R| \simeq |A_L|$ grow 
exponentially and independently of each other with the same growth rate; and their 
superposition causes SW-like oscillations with 
the large critical frequency as in Fig.~\ref{Fig:transient} a. Hence, ``almost'' 
mirror symmetric experimental 
setups that do not favor a particular TW propagation direction cause an initial 
growth phase with SW characteristics \cite{WK92}.

But then a competition between the two TW constituents sets in
when the advective nonlinearities have become sufficiently strong. They amplify
the mirror-symmetry breaking differences between $|A_R|$ and $|A_L|$ and cause 
the decay of the minority TW. Thus, the SW, which still has a large 
frequency, is transformed into a fast TW (Fig.~\ref{Fig:transient} b -h). This 
SW $\to$ TW transformation being
advection driven is spatiotemporally complicated, in particular for the
concentration field and it implies a dramatic redistribution of concentration by
advective ``rolling in'' of concentration (Fig.~\ref{Fig:transient} c -g).
It takes place within less than one vertical thermal diffusion time and it
starts roughly when the flow amplitudes of
the two constituent TWs have grown to about the phase velocity $v_p$:
First, concentration is advected up- and downwards in the form of plumes by the 
growing SW-like 
flow that reverts periodically its direction; the vertical concentration 
gradient being still as large as that of the quiescent conductive state.  
The SW $\to$ TW transformation is triggered by an advective wave breaking 
and wave toppling process of the crests and troughs of the concentration wave
(Fig.~\ref{Fig:transient} c -d), 
whereas the waves of $w$ and $\delta T$ do not undergo substantial structural 
changes. Advective nonlinearities have by now
become sufficiently strong to make the mirror-symmetry-breaking
differences between the original, left and right propagating TW constituents of 
the SW clearly visible. When the concentration wave crests (troughs) with a
high (low) alcohol content bend and topple, they 
are advectively ``rolled in''. 

The flow induced sequence of first deforming,
then bending, and finally rolling in the plume-like wave crests and troughs is 
associated with and driven by a growth of the spatial phase shift between 
velocity and concentration field from zero
to about $\lambda/4$ during the SW $\to$ TW transformation: In the SW
the spatial location of the nodes of $\delta C$ and $w$ coincide while their 
oscillations
 are shifted in time by about a quarter of an oscillation period. The value 
of this phase difference, $\varphi_C - \varphi_w \simeq \pi/2$, does not change 
 during the whole transition sequence
 but in the TW, it also implies a spatial shift of $\delta C$ and $w$ of $\lambda/4$.

In the early TW phase (Fig.~\ref{Fig:transient} h)
the lateral concentration difference between adjacent roll like regions is  still almost as large as the initial 
vertical concentration contrast in the conductive state. Thus, the frequency and 
the phase velocity of this emerging TW are still very large, i.e., 
not much smaller than the critical values. But then for the parameters 
of Fig.~\ref{Fig:transient} a long-term TW transient to a 
low-frequency, strongly nonlinear and strongly anharmonic relaxed TW state 
(Fig.~\ref{Fig:transient} i) sets in:
slow diffusion degrades and homogenizes the concentration striations, 
the spatial extension over which $\delta C$ is 
constant at the two alternating high and low levels increases, the plateau height
decreases, the width of boundary layers between these plateaus shrinks, the 
$\delta C$-wave profile becomes more and more trapezoidal, and the
alcohol surplus (deficiency) in the cold top (warm bottom) part of the fluid layer 
decreases, thereby reducing the overall vertical concentration difference between 
top and bottom. This longtime degradation of concentration gradients is reflected 
by a dramatic decrease of the mixing number $M$ and with it of
the frequency $\omega$ relative to the initial values --- the better the fluid
becomes mixed the smaller is $\omega$.

 \subsection{Standing waves}
 \label{Sec:SW}
 
 TW and LTW convection has been studied
experimentally and theoretically in detail 
\cite{review,Ahlers,Behringer,Kolodner,Platten,Steinberg,Surko,Yahata,Knobloch,JL02}. 
But little is known about 
nonlinear SW states beyond a weakly nonlinear analysis 
\cite{SZ93} that is restricted to the immediate vicinity of
the oscillatory threshold. It showed that SWs are unstable there, typically 
bifurcating backwards.

In order to obtain the solution branch, we stabilize the SW states by suppressing 
phase propagation
 (so TWs can not compete against SWs) and, if necessary, by
exerting a control procedure. The latter operates via the field amplitudes or the
heat current injected into the fluid  and the Rayleigh number in response to the 
instantaneous frequency $\omega$ and its temporal derivative $\Delta \omega /
\Delta t$ , respectively \cite{control}. In this way we
trace out the SW solution branch all the way from close to onset with large 
frequency to slowly oscillating SWs that eventually period-double into chaos.
 
\begin{figure}
\includegraphics[width=13cm]{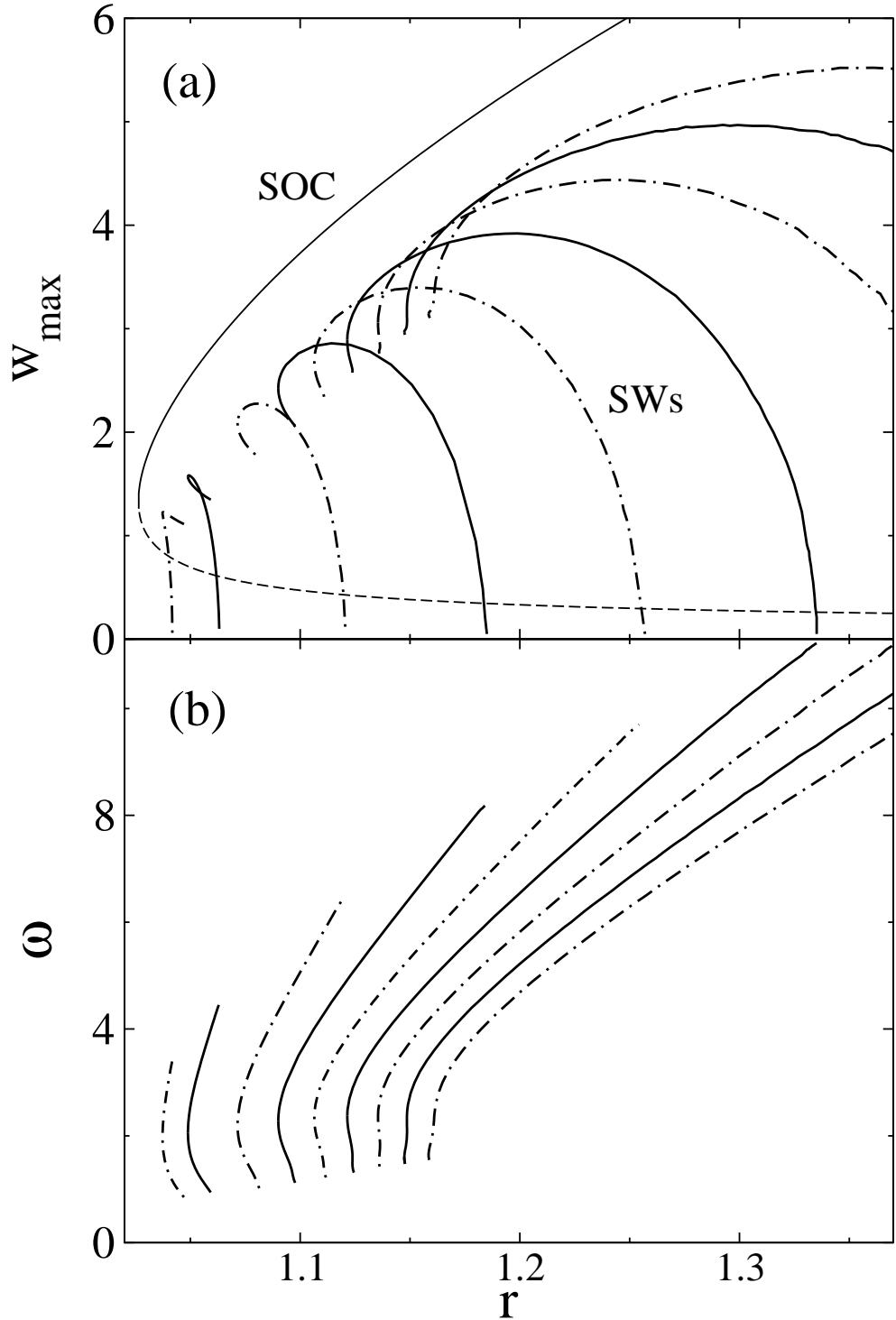}
\caption{Bifurcation properties of SWs for  $\psi$= -0.03, -0.05, -0.1, -0.15, 
-0.2, -0.25, -0.3, -0.35, -0.4 (from left to right) : (a) Maximal 
vertical velocity $w_{max}$. (b) Frequency $\omega$. Unstable SWs bifurcate subcritically out of the
quiescent conductive state [lower ends of the curves in (a); upper
ends in (b)] and undergo stability changes via saddle-node 
bifurcations. 
The SOC solution branch is shown for the sake of clarity only for $\psi$= -0.03.
SOC curves for the other $\psi$ are shifted slightly to the right.
Parameters are 
 $L =0.01$, $\sigma = 10$, and $\lambda = 2$.}
 \label{Fig:SWBifprops}
\end{figure}

Fig.~\ref{Fig:SWBifprops} shows  the bifurcation behavior of SWs and how it changes
with varying Soret coupling strength. The solution branch 
for the SOC is included for comparison only for $\psi=-0.03$.
The heating range in which SWs exist increases when $\psi$ becomes more 
negative since the oscillatory bifurcation threshold $r_{osc}$ is shifted 
stronger to higher $r$ than the SW saddle-node positions at $r_s^{SW}$ which marks the 
lower end of the  $r$-interval containing SWs. All these SWs bifurcate 
subcritically out of the conductive state as unstable solutions. They
become stable via saddle-node bifurcations when the phase-pinning condition is 
imposed. However, when 
this  condition is lifted completely then SWs decay by developing TW 
transients since any spatial phase difference between $\delta C$ and $w$ causes the extrema of
the latter to be "pulled" towards the solutally shifted buoyancy extrema.
Depending on $r$ these transients  either end in a nonlinear TW or SOC or the 
conductive state.

Moving along an SW branch the maximal 
vertical upflow velocity $w_{max}$ [Fig.~\ref{Fig:SWBifprops}(a)] does not
increase monotonically as in TWs and SOCs but rather has a maximum somewhat
below the respective SOC value before it drops again. On the other
hand, $\omega$ and $M$ decrease monotonically starting with the Hopf 
frequency $\omega_{H}$ [upper end of the curves in Fig.~\ref{Fig:SWBifprops}(b)]
 and $M=1$, respectively, at onset. This variation follows
 the universal scaling law $\omega/\omega_H \simeq M$, cf. Fig.~\ref{Fig:Mvsom}.

\begin{figure}
\includegraphics[width=12cm]{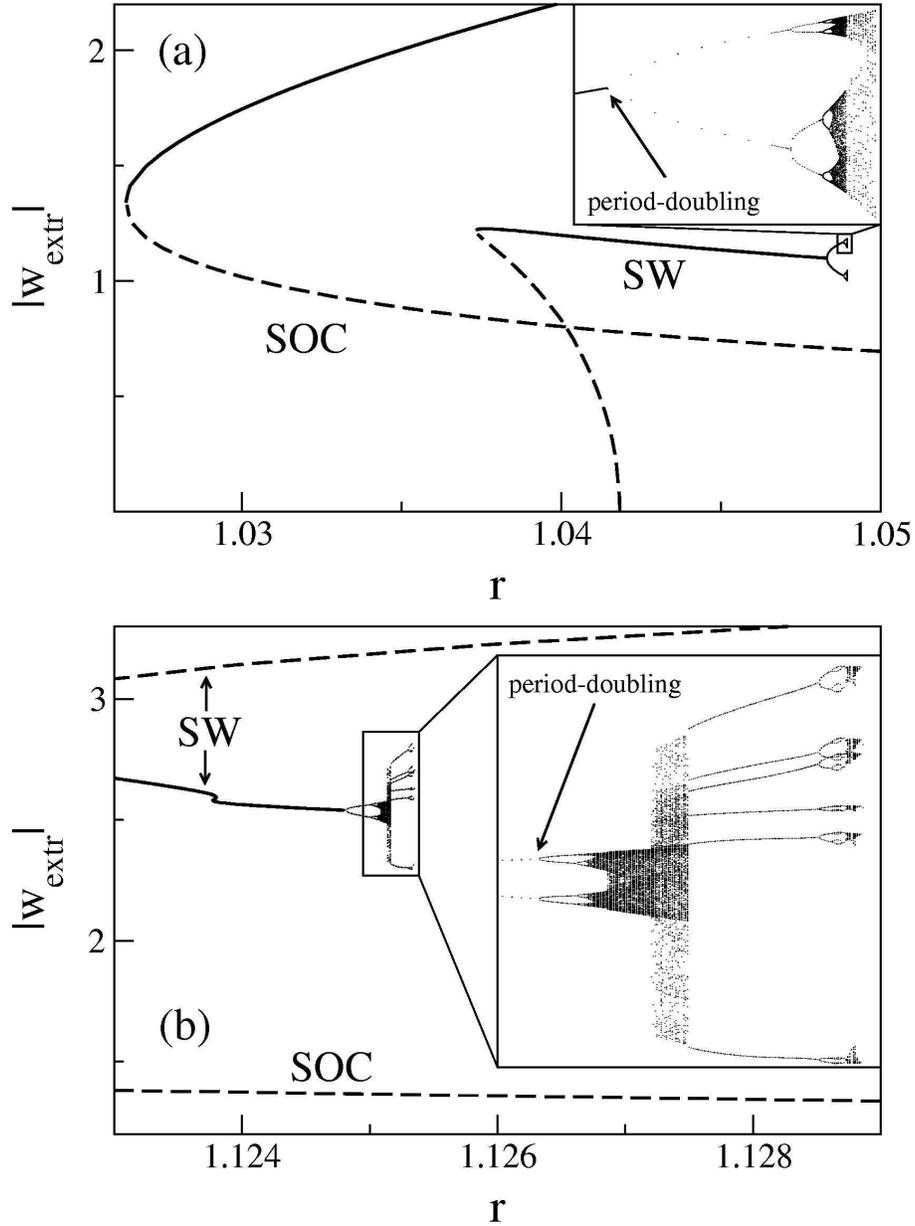}
\caption{Details of the SW bifurcation behavior for  $\psi=-0.03$ (a) and 
$\psi=-0.25$ (b).
The magnitude $|w_{extr}|$ of the extrema in the vertical flow is shown 
versus the reduced Rayleigh number $r$. 
The SWs bifurcate subcritically out of the quiescent conductive state at 
$r_{osc}$. When phase-pinning  conditions are applied they become stable 
(solid lines)  at saddle-nodes positions [lying outside the plot range of (b)]. 
When the  MTS is broken, the solid SW line 
splits into two since the magnitudes of the vertical flow extrema
occurring during one oscillation cycle become different [see, e.g., 
Fig.~\ref{Fig:SWPhasespace} (b) where the downflow at $x=0=z$ is more intense 
than the upflow]. 
This MTS-broken SW starts at the arrows to undergo  a period-doubling scenario
(cf. insets) leading to chaos.
Parameters are  $L =0.01$, $\sigma = 10$, and $\lambda = 2$.}
 \label{Fig:SWPdoubling}
\end{figure}

The typical bifurcation behaviour of two representative SWs of 
Fig.~\ref{Fig:SWBifprops} with $\psi$=-0.03 and -0.25  is
displayed in more detail in Fig.~\ref{Fig:SWPdoubling}.  Full (dashed) lines refer 
to stable (unstable)
branches. Note that the large-amplitude upper SOC solution branches 
[the one for $\psi$=-0.03 is shown
explicitly in Fig.~\ref{Fig:SWPdoubling}(a)] are stable down to their respective 
saddle-nodes when phase propagation is
suppressed so that TW solutions do not exist.
Then, the SW solutions, too, change their stability at  saddle-node bifurcation(s),
one for $\psi$=-0.03 
at $r^{SW}_{s}\approx1.0373$ and three for $\psi$=-0.25 [two of them can be seen in 
Fig.~\ref{Fig:SWPdoubling}(b) at  $r^{SW}_{s}\approx1.1238$]. 
With increasing  $r$ the flow amplitude of the stable SW  slightly decreases. Then  
the MTS breaks. Thereafter, for example, the downflow (upflow) extrema occurring 
in the SW oscillations, say, at $x$=0
($\pm \lambda/2)$ are more intense than the upflow (downflow) extrema. This is
reflected by the first splitting of the SW solution branches in 
Fig.~\ref{Fig:SWPdoubling}.
Consequently, the time averaged fields have now a net SOC-like structure 
with non zero mean downflow (upflow), say, at $x$=0 ($\pm\lambda/2)$.

\begin{figure}
\includegraphics[width=13cm]{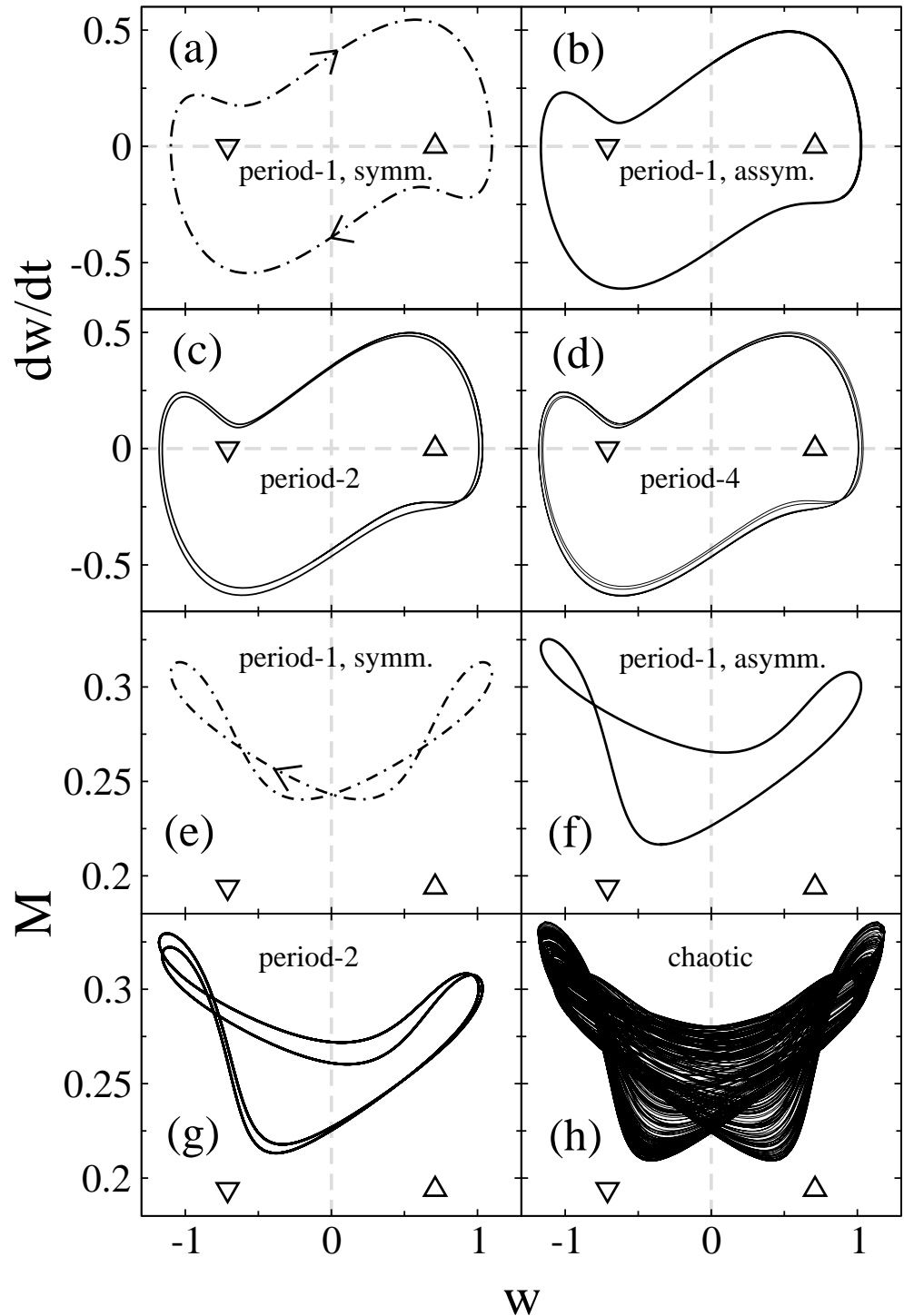}
\caption{MTS-breaking and period doubling in the phase space dynamics of SWs.
Shown are $\dot w$ and  the mixing number $M$ versus the vertical velocity $w$
at $x=0=z$ for $\psi$=-0.03. 
In (a)-(g)  the full (dash-dotted) lines refer to asymmetric (symmetric) SWs for which the MTS is (not yet)
broken. Period doubling is displayed in (b)-(d) and in (f)-(g). A chaotic trajectory is shown in (h).
Upwards and downwards 
pointing triangles indicate symmetry degenerate unstable SOC fixed 
points  with upflow and downflow, respectively, at $x=0$ that lie on the low-amplitude
SOC solution branch [dashed line in Fig.~\ref{Fig:SWPdoubling}(a)]. The flow velocity 
on the stable large-amplitude 
SOC solution branch [full line in Fig.~\ref{Fig:SWPdoubling}(a)] is $w \simeq 2.51$ for
the $r$-values displayed here.
Parameters are $L =0.01$, $\sigma = 10$, and $\lambda = 2$.}
 \label{Fig:SWPhasespace}
\end{figure}

In Fig.~\ref{Fig:SWPhasespace} we show for $\psi=-0.03$ how MTS breaking
changes the SW phase dynamics using $w,\dot w$ at the mid position $x=0=z$ and 
$M$ as characteristic local and global quantities, respectively. 
By definition $M$ oscillates with twice the SW frequency as
long as MTS holds (dash-dotted lines in Fig.~\ref{Fig:SWPhasespace}).
The particular MTS-broken SW orbits of 
Fig.~\ref{Fig:SWPhasespace} move closer to the low-amplitude fixed point 
on the lower SOC solution branch of Fig.~\ref{Fig:SWPdoubling}(a) with downflow at 
$x=0$, so the SW spends more time in the downflow phase at $x=0$ than in the upflow
phase. This can be seen more clearly in the temporal
oscillation profile of $w$ (Fig.~4 in \cite{MDL}) ; it develops a plateau close to 
the value of the corresponding SOC fixed point.

Then, at a certain $r$ depending on $\psi$, the first period-doubling 
occurs followed by further subsequent period-doublings that lead
to chaos (Fig.~\ref{Fig:SWPdoubling} and Fig.~\ref{Fig:SWPhasespace}).
For stronger Soret coupling, i.e. $\psi=-0.25$ [Fig.~\ref{Fig:SWPdoubling}(b)], we 
could also resolve 
a $r$-window with period-3 SW states and subsequent period doublings.

Beyond the last chaotic windows, we did not observe any stable SWs: heating 
above this threshold leads to the development of transients into a stable, 
large-amplitude SOC on the {\em upper} solution branch [e.g., with amplitude
$w_{SOC} \simeq 2.51$ in Fig.~\ref{Fig:SWPdoubling}(a), i.e., well outside the plot
range of Fig.~\ref{Fig:SWPhasespace}].

Thus, here the question arises whether -- and if so -- how and how long the 
heteroclinic orbits
connecting the two unstable symmetry degenerate SOC fixed points on the lower
solution branch with small convection amplitudes ($w_{SOC} \simeq 0.71$ in 
Fig.~\ref{Fig:SWPhasespace}) organize 
and restrain the dynamics of the periodic and chaotic SWs that switch 
between up- and downflow. 

 \section{Localized traveling waves}
 \label{Sec:LTW}

LTWs consist of wave trains of traveling convection rolls which are surrounded 
by quiescent fluid \cite{exp,LTW,Y91_NHY96,JL02}. There exist several attemps to 
model LTWs by weakly nonlinear small-amplitude expansions
around the convective onset \cite{vSH92,FT90} and modifications thereof 
\cite{DB90,BH90,Riecke92}.
But due to the strongly nonlinear characteristics of LTWs these models 
are aimed at some qualitative aspects of LTW states. 

Here we investigate stable LTWs at subcritical driving,
 $r<r_{osc}$, where the quiescent conductive surrounding is stable as well.
 
\begin{figure}
\includegraphics[clip=true,width=15cm,angle=0]{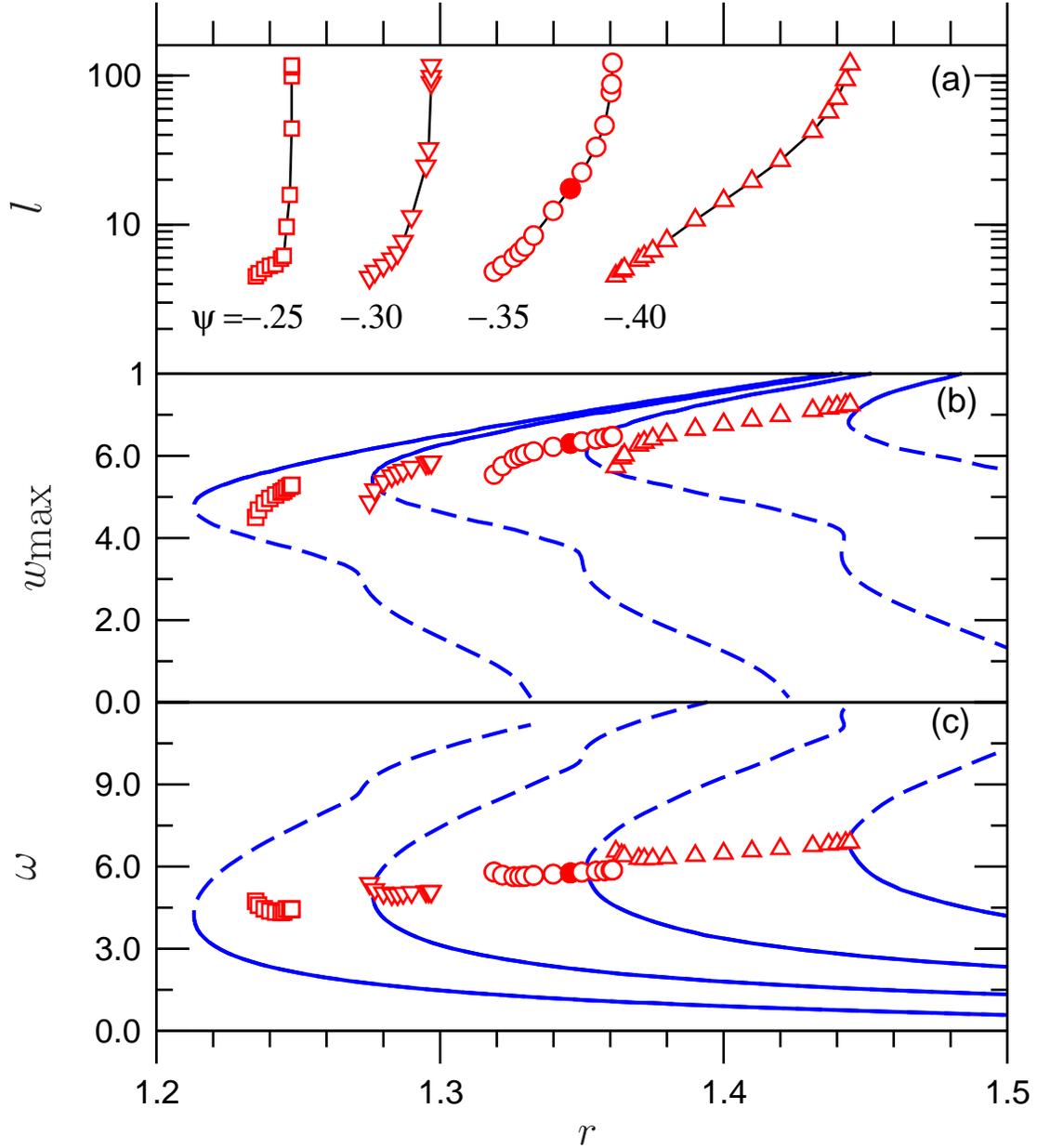}
\caption{Bifurcation properties of LTWs (symbols) and TWs (lines) for different
separation ratios $\psi$: (a) Full width $l$ of LTWs at half maximum of the 
envelope of the vertical velocity field $w$. (b) Maximal vertical flow 
velocity $w_{max}$. (c) Frequency (for LTWs in the frame comoving with the 
drift velocity $v_d$ of the respective LTW). Filled circles identify the LTW
whose structure is shown in Fig.~\ref{Fig:LTW}.
Lines in (b, c) denote TWs with 
saddle-node wave number $k_s^{TW} \simeq \pi$. 
  Unstable TWs (dashed lines; determined with a control
method) bifurcate subcritically with large Hopf frequency $\omega_H$ at 
$r_{osc}$ out of the conductive state and become stable (solid lines) at
the saddle-node $r_s^{TW}$ when lateral periodicity is imposed with
$\lambda=2\pi/k$. Parameters are $L = 0.01, \sigma =  10$. 
\label{Fig:BifTW_LTW}}
\end{figure}

\begin{figure*}
\includegraphics[width=12.0cm]{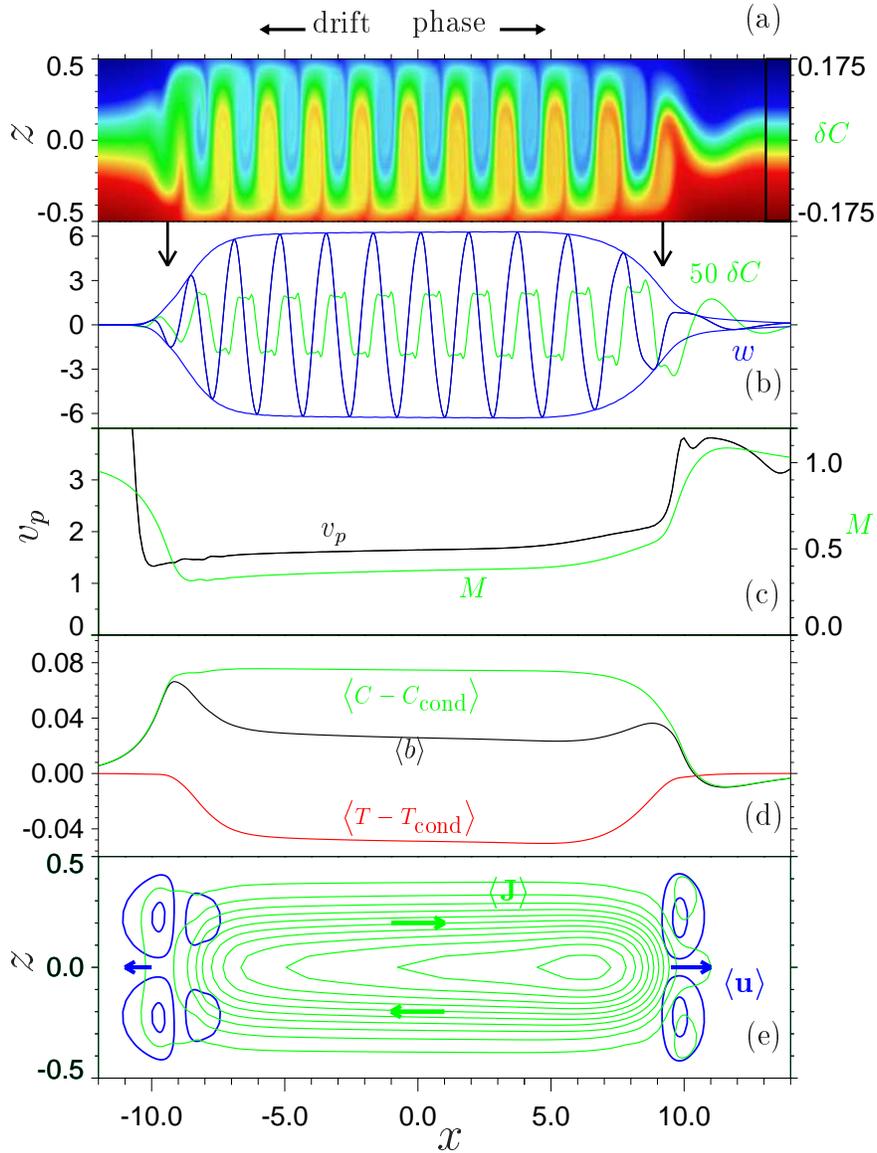}
 \caption{Broad LTW of $l$=17.4 identified by filled circles in 
 Fig.~\ref{Fig:BifTW_LTW}: (a) Snapshot of 
 concentration deviation $\delta C$ from the global mean value in
 a vertical cross section of the layer. (b) Snapshots of lateral wave
 profiles at midheight, $z$=0, of $\delta C$, vertical 
 velocity $w$, and its envelope. At the arrows $w_{max}=v_p$.
 (b) Mixing number $M(x)$, 
 [Eq.~(\ref{Eq:Mxdef})] and phase velocity 
 $v_p$ of nodes of $w(z=0)$ in the frame comoving with the small drift velocity $v_d$ of
 the LTW. The variation of 
 $\lambda(x)=2\pi\;v_p(x)/\omega$  is the same
 since the LTW frequency $\omega$ is a {\em global} constant.  
 (d) Time averaged deviations from the conductive state at $z$=-0.25 for 
 concentration, temperature, and their sum, $ b= T-T_{cond} + C-C_{cond} $,
 measuring the convective contribution to the buoyancy. 
 (e) Streamlines of time averaged velocity field $\langle \vec{u} \rangle$
 and concentration current  $\langle \vec{J}\rangle = 
 \langle \vec{u}\delta C - L\Nabla (\delta C -\psi \delta T)\rangle$. 
 $\langle \vec{u} \rangle$ results from $\langle b \rangle$ and
 affects $\langle \vec{J}\rangle$ via the contribution 
 $\langle \vec{u} \rangle \langle \delta C \rangle$.
 In the upper half of the layer positive $\delta C$ (alcohol surplus) is
 transported to the right. In the lower half of the layer negative 
 $\delta C$ (water surplus) is also transported to the right -- positive 
 $\delta C$ is transported there to the left as indicated by the arrow.
  Parameters are $L =0.01$, $\sigma = 10$, $\psi$=-0.35, and $r$=1.346. 
 \label{Fig:LTW}}
  \end{figure*}

The rolls grow out of this environment at the tail end of the
wave train, travel through the convective bulk of the LTW with increasing phase 
velocity $v_p(x)$ and wavelength $\lambda(x)$,
and decay at the head of the wave train (cf. Fig.~\ref{Fig:LTW}). These stable 
LTW states are uniquely selected. Their width $l$ (or number of rolls) is 
stationary and depends in an unique way on 
the control parameters (cf. Fig.~\ref{Fig:BifTW_LTW}).
The whole convective region drifts through the motionless state
with a drift velocity $v_d$ that can be positive or negative but that is small 
compared to the phase velocity \cite{JL02}.
Therefore, the frequency of a LTW is constant in a frame of reference that is
comoving with the LTW's drift.

To characterize the concentration variations in a LTW we use a
local mixing number $M(x)$ that is defined similarly to the 
case of extended states as the mean variance of $\delta C$. 
But here the lateral average is 
replaced by a time average at a fixed lateral position $x$,
\begin{equation}
\label{Eq:Mxdef}
M(x) = \sqrt{\langle \delta C^2\rangle / \langle \delta C^2_{\rm cond}\rangle} \,.
\end{equation}
So brackets denote here an average over $z$ and $t$ instead of $z$ and $x$ as in 
Eq.~(\ref{eq:Mdef}).
Fig.~\ref{Fig:LTW} (c) shows the variation of $M(x)$ over the convective region of a
typical long LTW. The relation between $M(x)$ and the local phase velocity $v_p(x)$
in the bulk of the LTW is the same as the relation between $M$ and $v_p=\omega/k$ 
for TWs.

The smallest stable localized pulses have for each $\psi$ a minimal width of 
about 5 rolls.
With increasing $r$  LTWs grow in width and amplitude and
$l$ diverges at a maximal driving $r_{max}$. Beyond this value,
LTWs of stationary width seem to be no longer possible. Thereafter, only transient
localized 
convective regions can occur which keep expanding between two fronts.
 
As one can see in Fig.~\ref{Fig:BifTW_LTW}
the existence interval of LTWs at fixed $\psi$ can be divided into two regimes:
In the regime of short pulses the frequency decreases with increasing $r$ and the amplitude
grows fast whereas in the regime of long LTWs the frequency increases with $r$ and 
the amplitude grows only moderately.

Short LTWs are dominated by a direct interaction between the growth part
and the decay part of the convection rolls.
Long LTWs are characterized by an extended bulk part where rolls travel with 
phase velocity $v_p(x)$ that increases from tail to head. This corresponds to an
increasing local wavelength $\lambda(x)=2\pi v_p(x)/\omega$ since 
the frequency $\omega$ is constant in the frame that is comoving
with $v_d$. The speeding up of $v_p(x)$ reflects 
the growing concentration contrast between adjacent rolls along the $x$-direction:
The minimal mixing number is located in the growth region of the tail where 
$v_p(x)$ and $\lambda(x)$ are minimal as well.
On their way from tail to head the rolls do not reach a stationary balance between
$\delta C$ injection from the different boundary layers
and advective mixing and diffusive homogenisation
of concentration differences on a constant level of small $M$.
Rather LTW rolls collapse at the head when $v_p$ has grown up to $w_{max}$ 
[right arrow in Fig.~\ref{Fig:LTW} (b)]. Thereafter concentration is discharged and
sustains a barrier of $\langle \delta C \rangle$ ahead of the decay part which 
stabilizes the conductive state there against invasion of convection.

Fig.~\ref{Fig:BifTW_LTW} (b) and (c) compare the maximal flow velocities and the 
frequencies of LTWs for several $\psi$ with the respective TW branch of  
$\lambda\simeq 2$ for which the saddle location $r_s^{TW}$ is the lowest one.
Note that the LTW values for $w_{max}$ and $\omega$ always lie near the saddle 
values of TWs. 
Although there exist no extended TWs below the shown TW saddle positions 
there exist stable LTWs ahead of these global TW saddles when $\psi$ is
sufficiently negative.
For example, for $\psi=-0.4$ almost all LTWs of finite width appear {\it below} the 
existence region for TWs as can be observed in Fig.~\ref{Fig:BifTW_LTW}.

The intriguing  existence of stable LTWs without coexisting TWs is ensured by a 
flow-induced
lateral concentration redistribution over its convective bulk.
Positive ``blue'' (negative ``red'') concentration deviation from the global mean 
is sucked from 
the top (bottom) boundary layer into right (left) turning rolls as soon as they
become
nonlinear during their growth [left arrow in Fig.~\ref{Fig:LTW} (b)].
Convection  in the bulk of a (long) LTW shows all the characteristics of a 
strongly nonlinear extended TW. Thus, as discussed in Sec.~\ref{Sec:Mixdiff},
positive (negative)
concentration is transported within the closed streamline regions predominantly 
in the upper (lower) half away from the tail towards the head.
This lateral concentration transport is reflected in the time averaged 
current
of $\delta C$ [cf. its streamlines in Fig.~\ref{Fig:LTW}(e)].
At the same time, mean concentration, $\delta C \simeq 0$, migrates mostly to the 
left along open velocity field streamlines 
that meander between the closed roll regions [green regions in
Fig.~\ref{Fig:LTW}(a)]. 
In this way a region with well mixed fluid, i.e., almost vanishing $\delta C$
is created and sustained under the tail.
 This strongly nonlinear large scale concentration redistribution that maintains 
the concentration locally at a homogeneous mean level under the trailing front
makes there the growth of convection possible: By reducing the stabilizing 
concentration gradients the driving buoyancy force is
locally increased there to levels which suffice to cause convection
in a well-mixed fluid. 

Fig.~\ref{Fig:LTW} (d)
shows that this effect causes the mean convectively generated $C$ profile 
to extend
significantly further into the conductive region than the mean convective 
temperature field.
Thus, the buoyancy $\langle b \rangle$ [cf. Fig.~\ref{Fig:LTW} (d)] is determined 
in the front regions predominantly
by the concentration field. This explains (i) the decrease of buoyancy below 
conduction levels
ahead of the decay part with the associated restabilization of conduction there 
and
(ii) the increase of $\langle b \rangle$ out of the conductive state at the tail 
and
its strong overshoot over the bulk enabling convection growth even for heating $r$
where no stable extended TWs are possible.

 \section{Conclusion}
 \label{Sec:conclusion}

Three oscillatory steady states of convection in binary fluid
mixtures like ethanol-water were investigated and compared numerically in a cross section
perpendicular to the roll axes:
Traveling waves, standing waves, and localized traveling waves.
In addition the supercritical growth of SWs out of perturbations and the 
SW $\to$ TW transformation was elucidated.

The full bifurcation behavior of TWs and SWs has been presented and their typical 
concentration field structures are shown in detail.
Both extended oscillatory periodic patterns bifurcate subcritically out of the conductive
state in a common Hopf bifurcation. The decrease of the phase velocity or the frequency along the solution
branches is connected to the decrease of the mean $\delta C$ contrasts in the system
due to advective mixing.
Thereby, the different symmetries and flow topologies of TWs and SWs cause a different
behavior in the strongly nonlinear regime.
However frequencies and mixing numbers are always related to each other via the universal
scale relation $\omega/\omega_H \simeq M$.

TWs are mirror-glide symmetric under $(x,z) \to (x+\lambda/2, -z)$. 
At high amplitudes they are characterized 
by the occurrence of closed streamlines in the frame comoving with the TW's phase
velocity. The regions of closed streamlines are displaced alternately towards the
upper and lower plates with their respective concentration boundary layers.
The latter feed selectively concentration only into the nearby closed streamline 
domains
where it is homogenized. This leads to the characteristic 
concentration distribution with homogeneous rolls of alternating signs in $\delta C$.
They are separated by meandering inner boundary layers of $\delta C \sim 0$ which are spatially
correlated with open streamlines of the velocity field. 
With increasing amplitude the concentration contrasts between the rolls decrease, the 
phase velocity slows down, and the TW ends at large amplitudes in a mirror symmetric 
SOC state.

SWs are always laterally mirror symmetric. The solution branch turns around at a saddle node
and becomes stable there under phase pinning conditions.
For strongly negative $\psi$ the stable part of the SW branch is directed towards lower amplitudes
-- in contrast to the TW case -- and develops further saddle nodes.
The complicated bifurcation behavior at low frequencies  seems to be due to an interaction
with the unstable low amplitude SOC solutions. There is a breaking of the mirror-timeshift
symmetry which is a prerequisite for the following period doubling cascade into chaos.
Without phase pinning the SWs develop TW transients which end either in a nonlinear TW or a SOC state
depending on the parameters.
This mechanism is similar to the symmetry breaking of a weakly nonlinear SW during
its transient growth at supercritical heating towards a nonlinear TW.
The transition is characterized by a dramatic advective reorganization of the $\delta C$ field and therefore
of the buoyancy field within a short time interval.

The only global symmetry operation for a LTW is the mirror-timeshift [$(x,z,t) \to (x,-z,t+\tau/2)$] 
in the frame
which is comoving with its small drift velocity $v_d$. The structual properties of a generic
long LTW were presented as well as the global bifurcation behavior with $r$ for different
$\psi$.
For strongly negative separation ratios LTWs exist monostably at low heatings $r$ where no
extended TWs are possible.
This is ensured by a flow-induced lateral large scale redistribution of concentration
over its TW dominated convective bulk:
Thereby, concentration is maintained under the tail of the wave train at a homogeneous mean level.
This increases the driving buoyancy forces there locally to levels which suffice to cause convection.



\clearpage


\begin{thebibliography}{999}
\bibitem{coupling}
For 5 weight percent of ethanol mixed into water at $T=20^{\circ} C$ the separation
ratio measuring the Soret coupling strength \cite{CH93} is 
$\psi \simeq -0.3$ \cite{KWM88}.
\bibitem{osc-inst} 
The oscillatory instability is not primarily due to concentration 
diffusion being small: even if the diffusive time scales for momentum, heat, 
and concentration are the same ($L = \sigma = 1$) there are oscillations.
They are caused by a 
sufficiently strong "off diagonal" Soret coupling between the degrees of 
freedom of concentration and temperature and their feedback into the 
buoyancy.
\bibitem{reviewCF}
For a review see \cite{PL84,CH93}. Further references may be found, 
e.g., in \cite{BLKS95I,LBBFHJ98}. See also the more recent publications
\cite{KKS94a,KGW90-KFK99,TPC96,NHY97b,LCA99,AS01,BNMK01} of some of the many 
groups that have investigated convection in binary mixtures.
\bibitem{control}
We drive the system towards the unstable limit cycle (ULC) with frequency 
$\omega_\ast$ that is located on the subcritical SW solution branch as follows:
To reduce deviations with growing (decreasing) convection 
amplitude $A$ that decrease (increase) $\omega$ at fixed $r$ we set 
$A_{n+1}=A_n [1 + W_A(\omega - \omega_\ast)]$. Furthermore, when 
$\Delta \omega / \Delta t$ is negative (positive) we move towards  
the SW solution branch by decreasing (increasing) $r$ according to 
$r_{n+1}=r_n (1 + W_r \Delta \omega / \Delta t)$. The control depends on the
actual values of $\omega - \omega_\ast$ and $\Delta \omega / \Delta t$ and
vanishes only at the ULC with the preselected frequency $\omega_\ast$. The 
optimal weight factors $W_A$ and $W_r$ depend on $\omega_\ast$.
\end{thebibliography}
\end{document}